\documentclass[12pt]{article}
\usepackage{amsmath}
\usepackage{graphicx}
\usepackage{enumerate}
\usepackage{setspace}
\usepackage{natbib}
\usepackage{amsthm,amsfonts}
\usepackage[usenames,dvipsnames]{xcolor}

\newcommand{\blind}{1} 

\newtheorem{defi}{Definition}

\newcommand{\abs}[1]{\left|#1\right|}

\DeclareMathOperator{\N}{\mathcal{N}}
\DeclareMathOperator{\FLP}{\mathcal{FLP}}
\DeclareMathOperator{\NFLP}{\mathcal{N-FLP}}
\DeclareMathOperator{\Bin}{\mathcal{B}}
\DeclareMathOperator{\re}{\mathbb{R}}
\DeclareMathOperator{\xb}{\mathbf{x}}
\DeclareMathOperator{\yb}{\mathbf{y}}
\DeclareMathOperator{\vb}{\mathbf{v}}
\DeclareMathOperator{\simdist}{\stackrel{\mathcal{L}}{\sim}}
\DeclareMathOperator{\approxdist}{\stackrel{\mathcal{L}}{\approx}}
\DeclareMathOperator{\diag}{diag}
\DeclareMathOperator{\Var}{Var}
\DeclareMathOperator{\midd}{\hspace{.05mm}|\hspace{.05mm}}
\DeclareMathOperator{\I}{\boldsymbol I}
\DeclareMathOperator{\betab}{\boldsymbol\beta}
\DeclareMathOperator{\Dpi}{\boldsymbol D_{\hat{\boldsymbol\pi}}}
\DeclareMathOperator{\Dv}{\boldsymbol D_\mathbf{v}}

\begin{document}

\if1\blind
{
  \title{\bf Efficient and Robust Estimation of Linear Regression with Normal Errors}
  \author{Alain Desgagn\'{e} \\ D\'{e}partement de math\'{e}matiques, Universit\'{e} du Qu\'{e}bec \`{a} Montr\'{e}al,\\ Montr\'{e}al, Canada, desgagne.alain@uqam.ca}
  \maketitle
} \fi

\if0\blind
{
  \begin{center}
    {\LARGE\bf  Efficient and Robust Estimation of Linear \\\vspace{5mm}    Regression with Normal Errors}
  \end{center}
} \fi

\begin{abstract}
Linear regression with normally distributed errors -- including particular cases such as ANOVA, Student's $t$-test or location-scale
inference -- is a widely used statistical procedure. In this case the ordinary least squares estimator possesses
remarkable properties but is very sensitive to outliers.
Several robust alternatives have been proposed, but there is still significant room for improvement.
This paper thus proposes an original method of estimation that offers the best efficiency simultaneously in the absence and the presence of outliers,
 both for the estimation of the regression coefficients and the scale parameter.
The approach first consists in broadening the normal assumption of the errors to a mixture of the normal and the
filtered-log-Pareto (FLP), an original distribution designed to represent the outliers. The expectation–maximization (EM) algorithm is then adapted
and we obtain the N-FLP estimators of the regression coefficients, the scale parameter
 and the proportion of outliers, along with probabilities of each observation being an outlier.
The performance of the N-FLP estimators is compared with the best alternatives in an extensive Monte Carlo simulation.
The paper demonstrates that this method of estimation can also be used for a complete robust inference,
including confidence intervals, hypothesis testing and model selection.
\end{abstract}


\section{Introduction}
\label{sec:introduction}

Linear regression with normally distributed errors -- including particular cases such as ANOVA, Student's $t$-test or location-scale
inference -- is a widely used statistical procedure.
In this case, ordinary least squares (OLS), or equivalently maximum likelihood estimation, possesses
remarkable properties such as minimum variance among unbiased estimates. However it is well known that the resulting inference is very sensitive to outliers,
which are defined in this paper as observations with ``extreme'' errors that conflict with the normal assumption.

Several robust methods of estimation have been proposed in the literature to address this situation, e.g.,
the M (\cite{huber1973robust}), S (\cite{rousseeuw1984robust}), least median of squares (LMS,  \cite{Rousseeuw1984LMS}),
least trimmed squares (LTS, \cite{Rousseeuw1985LTS}),  MM (\cite{Yohai1987MM}),
robust and efficient weighted least squares (REWLS, \cite{GerviniYohai2002rewlse}) estimators, to name the most popular.
In their review, \cite{yuyao2017review} concluded that the REWLS and MM estimators
perform the best by far in the estimation of the regression coefficients, achieving high efficiency simultaneously in the absence and the presence of outliers.

However, the REWLS and MM estimators are perfectible in several ways, which is what we propose to do in this paper with the introduction
of an original method of estimation. First, these robust methods of estimation focus on regression coefficients; the estimation
of the scale parameter is overlooked to a certain extent. This parameter nonetheless plays a crucial role in statistical inference such as confidence intervals,
hypothesis testing and model selection. To address that, we propose a scale estimator that performs much better than the existing alternatives, both in the absence and the presence of outliers.
Second, we offer a significant improvement in efficiency in the estimation of the regression coefficients, with or without outliers.
 Third, one downside of robust methods for practitioners mainly familiar with OLS is that they obtain different results when the sample is free of outliers.
Among the estimators mentioned above, only the REWLS can generate identical results to OLS for small fractions of uncontaminated samples, solely for the estimation of the regression coefficients. We greatly improve this aspect by offering a method with the same nice feature, for the estimation of both the regression coefficients and the scale parameter,
for a large majority of samples without outliers (see Section~\ref{sec:MonteCarlo} for more details).

To achieve our objective, we assume that the distribution of the errors is
a mixture of the normal with another distribution  representing the outliers, instead of
the pure normal. Our first key contribution is thus to design a specific parametric distribution for the outliers,
that we name ‘filtered-log-Pareto’ (FLP) distribution, defined only on the tails in the spirit of the outlier region defined by \cite{DaviesGather1993}.
The resulting mixture is named the FLP-contaminated normal or simply the N-FLP distribution. This original mixture
is designed such that the outlier region and the tail's behavior are set automatically, based on the proportion of outliers given by the mixture weights.
The N-FLP density is thus a normal density downweighted in the central part with, in return,
its tails thickened to accommodate for possibly more extreme values than expected under the normal model.

Our second original contribution is to adapt the well-known expectation-maximization (EM) algorithm for the estimation of the N-FLP mixture.
We thus obtain explicit and interpretable expressions that we name ``N-FLP estimators'', for the mixture weights
and for both the regression coefficients and scale estimators in the same forms as the weighted least squares.
This transparency is certainly a desirable feature for practitioners who view statistical tools as decision aids. Further,
we obtain a probability, for each observation, of arising from the outlier component, which makes outlier detection very easy and efficient.

The remainder of the paper is structured as follows. In Section~\ref{sec:FLP-contaminated-normal}, the N-FLP mixture is introduced.
In Section~\ref{sec:linear-regression-model}, we describe the adapted EM algorithm that leads to the N-FLP estimators
of the linear regression model. A detailed example is given in Section~\ref{sec:example} where we address robust inference such as
confidence intervals,  hypothesis testing and model selection, through different specific models such as location-scale, simple and multiple linear regression, ANOVA and
Student's $t$-test. In Section~\ref{sec:MonteCarlo} the performance of our
approach is compared with the best and the most popular estimators (as listed above) through an extensive Monte Carlo simulation.
Section~\ref{sec:conclusion} concludes the paper.

\section{FLP-Contaminated Normal Distribution}
\label{sec:FLP-contaminated-normal}

The first step of our robust approach consists in assuming that the errors of the regression model have a FLP-contaminated normal distribution, also named N-FLP mixture.

\begin{defi}
\label{def:FLP-contaminated-normal}
A random variable $Y$ is said to have a FLP-contaminated normal distribution, given by the mixture
\begin{equation*}
  \NFLP(\omega,\mu,\sigma)=\omega \N(\mu,\sigma^2) + (1-\omega)\FLP(\omega,\mu,\sigma),
\end{equation*}
if the contaminating distribution is defined as a filtered-log-Pareto (FLP) distribution, which in turn
has a density defined by
\begin{align*}
f_{\FLP}(y\mid\omega, \mu, \sigma)=\left\{
  \begin{array}{ccc}
     0  & \text{if} & |z|\le \tau \text{ or }\omega=1, \\
     (1-\omega)^{-1}\omega\sigma^{-1}\left[\varphi(\tau)\frac{\tau}{|z|}
     \left(\frac{\log \tau}{\log |z|}\right)^{\lambda+1}-\varphi(z)\right]
     & \text{if} & |z|> \tau \text{ and }\omega<1, \\
  \end{array}
\right.
\end{align*}
where  $z=\sigma^{-1}(y-\mu)\in\re$, $0<\omega\le 1$ is the mixture weight for the normal component, $\mu\in\re$ is a location parameter and $\sigma>0$ is a scale parameter.

Furthermore, the tail's behavior is controlled by $\lambda=2(1-\omega\rho)^{-1}\omega \varphi(\tau)\tau\log\tau>0$,
with $\rho=2\Phi(\tau)-1$, where $\varphi(\cdot)$ and $\Phi(\cdot)$ are respectively the probability and cumulative distribution functions of a standard Gaussian.
Finally, the outlier region is controlled by $\tau>1.69901$ (precisely $\tau>1$ and $(\tau ^ 2-1)\log\tau > 1$), defined as
\begin{equation*}
  \tau=g^{-1}(\omega) \text{ with }g(\tau)=\left(\rho + \frac{2\varphi(\tau) \tau  \log\tau}{(\tau ^ 2-1)\log\tau - 1} \right)^{-1}.
\end{equation*}
\end{defi}

\begin{figure}[t]
\begin{center}
\begin{tabular}{cc}
\includegraphics[width=0.49\textwidth]{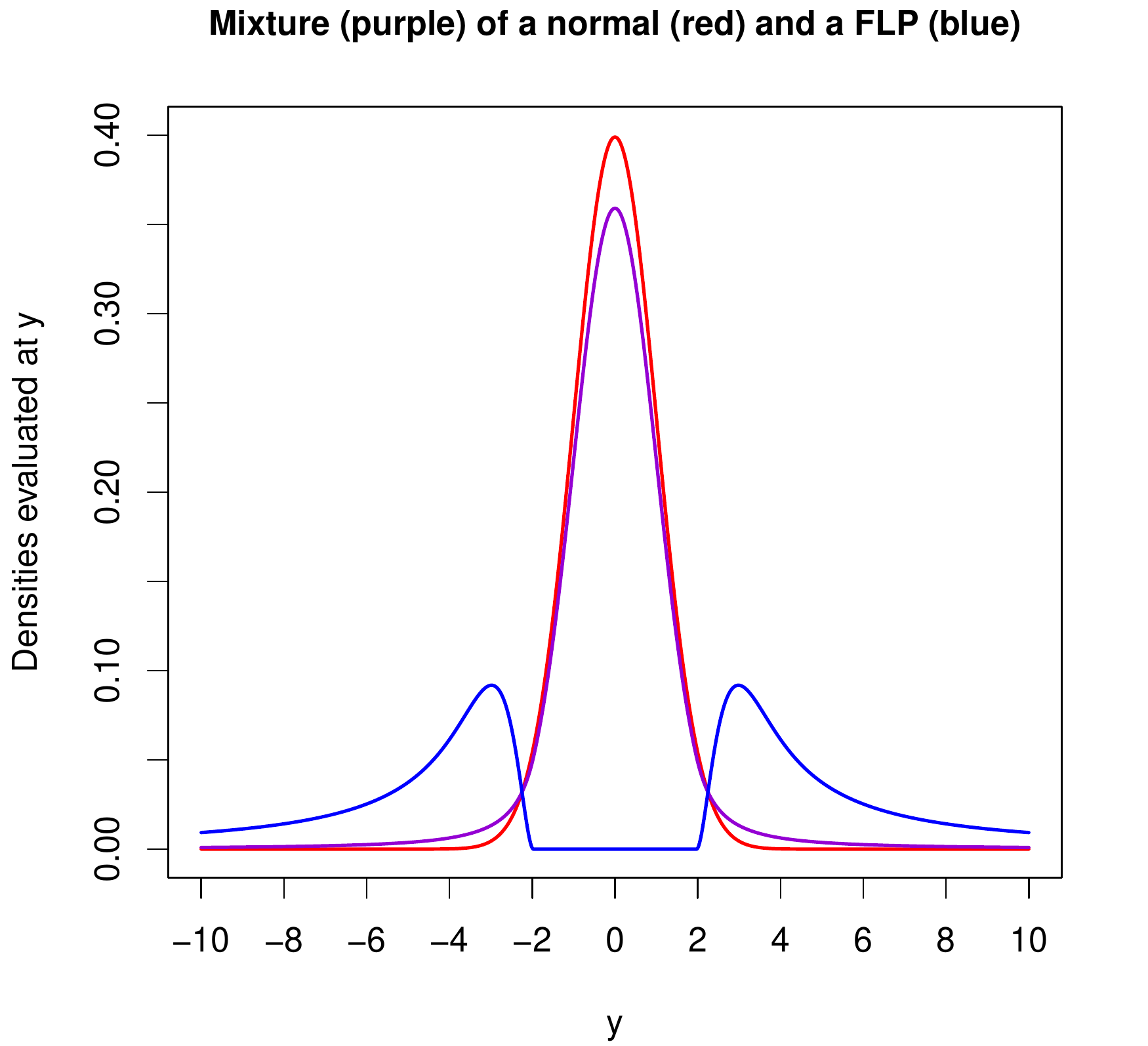}
\end{tabular}
\end{center}
\caption{The $\NFLP(\omega,\mu, \sigma)$ density  (purple) as a mixture of the $\N(\mu, \sigma)$ (red) and the $\FLP(\omega,\mu, \sigma)$ (blue) for specific values of
   $\omega=0.90$, $\mu=0$ and $\sigma=1$.}
\label{fig:mixture}
\end{figure}

The density of the $\NFLP(\omega,\mu,\sigma)$ for specific values of $\omega=0.90$, $\mu=0$ and $\sigma=1$ is illustrated in Figure~\ref{fig:mixture},
along with its mixture components given by the $\N(\mu,\sigma)$ and the $\FLP(\omega,\mu,\sigma)$,  all three being continuous on the real line and symmetrical with respect to the location $\mu$.
The density of the mixture $\NFLP(\omega,\mu,\sigma)$, given explicitly by
\begin{align*}
  f_{\NFLP}(y\mid\omega, \mu, \sigma)&=\omega f_{\N}(y\mid \mu,\sigma) + (1-\omega)f_{\FLP}(y\mid\omega, \mu, \sigma)\\
  &=\left\{
  \begin{array}{ccc}
     \omega\sigma^{-1} \varphi(z)  & \text{ if } & |z|\le \tau \text{ or }\omega=1, \\
     \omega\sigma^{-1}\varphi(\tau)\frac{\tau}{|z|}
     \left(\frac{\log \tau}{\log |z|}\right)^{\lambda+1}
     & \text{ if } & |z|> \tau \text{ and }\omega<1, \\
  \end{array}
   \right.
\end{align*}
is a normal downweighted by $0<\omega\le 1$ in the central part.
In return, its tails -- defined as the outlier region $|z|> \tau$ -- are raised everywhere to accommodate for possibly more extreme values than expected under the normal model.
The $\NFLP(\omega,\mu,\sigma)$ distribution should in fact be interpreted as the mixture $\omega\N(\mu,\sigma^2) + (1-\omega) \FLP(\omega_0,\mu_0,\sigma_0)$, where
$(\omega_0,\mu_0,\sigma_0)$  form a distinct set of parameters with different meanings but are purposely set to the mixture weight, location
and scale parameters of the normal component given by $(\omega,\mu,\sigma)$.
We observe that the $\FLP(\omega,\mu,\sigma)$ is well defined as a probability distribution because
the equation of $\lambda$ ensures that its density integrates to 1 and the equation of $\tau$ ensures that it is positive. The proof
is given in Appendix~A in the supplemental material.

An important characteristic of the $\NFLP(\omega,\mu,\sigma)$ distribution is its capacity to adjust its shape to the number of outliers.
Indeed, the threshold $\tau>1.69901$  that defines the outlier region $|z|>\tau$ and the tails' decay determined by $\lambda>0$
are set automatically as a function of $\omega$, the expected proportion of normal observations. An increase in $\omega$ (meaning fewer expected outliers) results directly
in an increase in $\tau$ and $\lambda$, which translates into a narrower outlier region and lighter tails.
For example, if $\omega=0.90$, we obtain $\tau = 1.9709$ and $\lambda=0.9571$.
If $\omega=0.95$, we obtain $\tau = 2.1045$ and $\lambda=1.5512$.
At the limits, as $\omega$ approaches 0 or 1, the value of $\tau$ approaches 1.69901 or infinity and the value
of $\lambda$ approaches 0 or infinity.
If $\omega=1$, the model is simply the Gaussian distribution, with no contaminating component.

Further remarks can be made on the FLP-contaminated normal model.
Given that $\rho$ represents the probability of the central part under the $\N(\mu,\sigma)$ model,
the probability $\omega$ of generating a normal observation can be split into
probabilities $\omega\rho$ and $\omega(1-\rho)$  that an observation comes
from the normal truncated respectively on the central part $|z|\le \tau$ and the outlier region $|z|> \tau$.
An observation is thus generated from the tails $|z|> \tau$ with probability $1-\omega\rho$,
either from the truncated $\N(\mu,\sigma)$ with probability $\omega(1-\rho)$ or from the $\FLP(\omega,\mu,\sigma)$ with probability $1-\omega$.
Therefore, given that an observation has been generated from the tails, the probabilities
are $(1-\omega\rho)^{-1}\omega(1-\rho)$ that it was from the (truncated) normal component and $(1-\omega\rho)^{-1}(1-\omega)$ that
it was from the contaminating component $\FLP(\omega,\mu,\sigma)$, which defines a mixture distribution. It turns out that this mixture can be interpreted as a (double) log-Pareto distribution. In that sense, the $\FLP$ represents the non-normal component of the log-Pareto mixture, hence the name ``filtered-log-Pareto'' distribution.

As an alternative to the super heavy-tailed log-Pareto, we initially considered other tail decays, namely the heavy-tailed Pareto or a light-tailed distribution such as
the shifted exponential. However, their performance in the presence of a large proportion of outliers was negatively impacted, mainly
due to an oversized outlier region in order to compensate for their lighter tails. In contrast,
the largest theoretical outlier region (which happens when $\omega\rightarrow 0$) for the N-FLP model is
 given by $|z|>1.69901$, which leaves a fully acceptable mass of $\rho=0.91$ in the central part.

\section{Robust Estimation of the Linear Regression Model}
\label{sec:linear-regression-model}

The linear regression model is first given in Section~\ref{sec:regression-model}.
The N-FLP estimators are presented in Section~\ref{sec:N-FLP-estimators}, where our approach of adapted EM algorithm is described.
We go a step further in Section~\ref{sec:complete-inference} where different aspects of a complete robust inference such
as confidence intervals, hypothesis testing and model selection are discussed. Finally, the topic of outlier identification
is covered in Section~\ref{sec:outlier-detection}.

\subsection{Linear Regression Model}
\label{sec:regression-model}

We are given a random sample $(\xb_i, y_i)$, $i=1,\ldots,n$, where the response variable $y_i\in\re$ is related to $\xb_i\in\re^p$, a vector of $p\ge 1$ explanatory
 variables (the first element is set to 1 for a model with an intercept),  through the linear regression model
\begin{equation*}
\label{eqn:model-reg}
    y_i=\xb_i^T \betab + \varepsilon_i.
\end{equation*}
The vector of regression coefficients is $\betab:=(\beta_1,\ldots,\beta_p)^T\in\re^p$ and the error terms $\varepsilon_1,\ldots,\varepsilon_n$ are i.i.d. random variables
with the normal distribution: $\varepsilon_i\simdist \N(0,\sigma^2)$, where $\sigma>0$ is a scale parameter.
This is the classical linear regression model with Gaussian errors.

\subsection{N-FLP Estimators}
\label{sec:N-FLP-estimators}

Our robust approach first consists in broadening the normal assumption of the errors to the FLP-contaminated normal distribution, given by
\begin{equation*}
  \varepsilon_i\simdist \NFLP(\omega, 0,\sigma)=\omega\N(0,\sigma^2) + (1-\omega) \FLP(\omega,0,\sigma),
\end{equation*}
where $\sigma>0$ is the scale parameter and $0<\omega\le 1$ is the mixture weight for the normal component.
If $\omega=1$, we obtain the classical linear regression model with Gaussian errors.
Note that the specific choice of the FLP distribution for the outlier component is intended in fact
to represent any distributions of the outliers.
We first present the N-FLP estimators of the parameters $\omega$, $\betab$ and $\sigma$ and then we discuss how they have been generated using an adapted EM algorithm.
\begin{defi}
\label{def:estimators}
The N-FLP estimators are given by
\begin{equation*}
  \hat{\omega}=\frac{1}{n}\sum_{i=1}^n \hat{\pi}_i, \,\,\,\,\,\,
  \hat{\betab}=(\xb^T \Dpi \xb)^{-1}\xb^T \Dpi\yb \,\,\,\,\text{ and }\,\,\,\,\hat{\sigma}^2=\frac{1}{(\sum_{i=1}^n \hat{\pi}_i-p)}\sum_{i=1}^n \hat{\pi}_i\,(y_i-\xb^T_i \hat{\betab})^2,
\end{equation*}
\begin{equation*}
  \text{with }
  \hat{\pi}_i\equiv\pi_{\hat{\omega}}(r_i)=\frac{\hat{\omega} f_{\N}(y_i\mid \xb_i^T \hat{\betab}, \hat{\sigma})} {f_{\NFLP}(y_i\mid \hat{\omega}, \xb_i^T \hat{\betab}, \hat{\sigma})}=
   \frac{\hat{\omega} f_{\N}(r_i\mid 0, 1)} {f_{\NFLP}(r_i\mid \hat{\omega}, 0, 1)},\,\,\text{ where }\,\,r_i=\frac{y_i- \xb_i^T \hat{\betab}}{\hat{\sigma}},
\end{equation*}
$\Dpi:=\diag(\hat{\pi}_1,\ldots,\hat{\pi}_n)$, $\xb:=(\xb_1,\ldots, \xb_n)^T$ and $\yb:=(y_1,\ldots,y_n)^T$.
\end{defi}

In particular, we obtain the OLS estimates when $\hat{\omega}=1$. Note that $\hat{\omega}=1\Leftrightarrow \hat{\pi}_1,\ldots,\hat{\pi}_n=1$
(or equivalently $\Dpi=\I_n$, where $\I_n$ is the identity matrix of size $n$).
The estimates are found following an iterative process that alternates between the simultaneous computation of $\hat{\omega}$, $\hat{\betab}$, $\hat{\sigma}$
and that of $\Dpi$, in the spirit of the EM algorithm. Initial values of $\hat{\omega}$, $\hat{\betab}$ and $\hat{\sigma}$ must therefore be provided by the user.
The process is stopped when convergence is reached for each parameter $\omega$, $\betab$ and $\sigma$,
that is when the difference between estimators of two successive iterations is below a chosen threshold such as $10^{-9}$.

Let us see how this procedure has been constructed. The first method of estimation that comes to mind is maximum likelihood estimation (MLE)
for the model $y_i\simdist\omega\N(\xb_i^T \betab,\sigma^2) + (1-\omega) \FLP(\omega,\xb_i^T \betab,\sigma)$, which can also be achieved by the EM algorithm.
The main drawback of this approach in our context is that the FLP component depends on the parameters $\betab$ and $\sigma$ while it is reasonable to assume that
only the normal observations should contain information about these parameters. The FLP component also depends on $\omega$, which should be dedicated exclusively
to the expected proportion of normal observations. As an alternative, we could use a distinct set of parameters $\omega_0, \betab_0, \sigma_0$
for the contaminating component and find MLE with the EM algorithm. In doing so, however, the number of parameters doubles and the
resulting N-FLP density loses its smoothness and symmetry.

Our solution to address these issues lies somewhere in between the two solutions described above. We assume that the model is
$y_i\simdist\omega\N(\xb_i^T \betab,\sigma^2) + (1-\omega) \FLP(\omega_0,\xb_i^T \betab_0,\sigma_0)$ and we use the EM algorithm
on the normal component to estimate $\omega$, $\betab$ and $\sigma$, which produces the estimators given in Definition~\ref{def:estimators}
(with a bias correction for $\hat{\sigma}^2$). The EM algorithm is, however, adapted for the estimation of the FLP component.
We proceed otherwise by setting $\hat{\omega}_0=\hat{\omega}, \hat{\betab}_0=\hat{\betab}, \hat{\sigma}_0=\hat{\sigma}$.
The E step of the algorithm follows normally and we obtain the equation of $\hat{\pi}_i$.
Note that if the sample is theoretically generated from the $\NFLP(\omega,\xb_i^T \betab,\sigma)$ model,
and especially the $\N(\xb_i^T \betab,\sigma)$, the first two
approaches of estimation are asymptotically equivalent given the properties of the MLE. It can be shown that this is also the case for our in-between approach,
which thus shares the same asymptotic properties, especially when the errors are normal.

It is always preferable to run multiple searches with different initial values.
We found in our Monte Carlo simulations that 10 runs with initial values of $\hat{\betab},\hat{\sigma}$  set to the REWLS estimates
(in adding random errors around $\hat{\betab}$) and $\hat{\omega}$ set to 0.80 work very well. The iterative process always converges and the number of solutions is typically one or two, rarely larger than three.
If we set $\hat{\omega}=1$ at the initial stage, convergence is immediate for any sample and we thus always obtain the OLS estimates as a base solution.
When this is the only solution proposed by multiple searches, the N-FLP estimators are the OLS estimates, and it can be concluded that there are no outliers in the sample.

If solutions other than the non-robust OLS with $\hat{\omega}=1$ are found, a choice has to be made.
Under the EM algorithm, the solution with the maximum likelihood would naturally be chosen. However it is not appropriate for our adapted EM algorithm
because the fitting is done essentially on the  normal observations and their estimated number $\hat{\omega}n=\sum_{i=1}^n \hat{\pi}_i$ differs for each solution proposed by the algorithm.
Therefore we propose to simply choose the most robust solution, that with the minimum value of $\hat{\omega}$,
provided it is larger than 0.50, to avoid the rejection of a majority of observations and to obtain a breakdown point (BP) of approximately 50\%.
This is the automatic procedure we use in the Monte Carlo simulation in Section~\ref{sec:MonteCarlo}. We found that this strategy
has a very small impact on efficiency in the absence of outliers, but in return, the gain in the presence of outliers is remarkable.

Naturally, practitioners can decide to analyze each of the proposed solutions and to use their best judgment to make a final choice. Or one
could decide to set the minimum value of $\hat{\omega}$ at a larger value such as 0.70, and consequently reduce the BP to approximately 30\%,
which can be well advised for small sample sizes such as $n=10$.
In any case, our algorithm provides all the credible solutions in full transparency. It can be used in automatic mode, which is recommended
for most users because of its simplicity and efficiency, or in manual mode.

\subsection{Robust Inference Using N-FLP Estimators}
\label{sec:complete-inference}

Once we obtained the N-FLP estimates of $\hat{\omega}$, $\hat{\betab}$, $\hat{\sigma}$ along with the probabilities $\hat{\pi}_1,\ldots,\hat{\pi}_n$,
we can go a step further for a complete robust inference with confidence intervals, hypothesis testing and model selection.
To this end, let $\vb=(v_1,\ldots,v_n)^T$ be an unobserved vector of $n$ independent latent binomial variables defined as $v_i=1$ if the observation
$(\xb_i, y_i)$ originates from the normal component or $v_i=0$ if it comes from the FLP contaminating distribution, with
$v_i\midd \omega \simdist \Bin(1, \omega)$.  We first assume that $\vb$ is known and proceed with the inference
using only the normal observations. The classical results adjusted by the latent variables are then found.
In a second step, the latent variables $v_i$ are estimated by $\hat{\pi}_i$. Fundamentally, this is how  $\hat{\omega}$, $\hat{\betab}$ and $\hat{\sigma}$
have been found because the OLS estimators on the normal observations are
$\hat{\betab}\midd \vb=(\xb^T \Dv \xb)^{-1}\xb^T \Dv \yb$ and
$\hat{\sigma}^2\midd \vb=(\sum_{i=1}^n v_i - p)^{-1}\sum_{i=1}^n v_i(y_i-\xb^T_i \hat{\betab})^2$,
where $\Dv=\diag(v_1,\ldots,v_n)$, and the MLE of $\omega$ is $\hat{\omega}\midd \vb =n^{-1}\sum_{i=1}^n v_i$.

For example, the variance of $\hat{\betab}$ when $\vb$ is known is given by
\begin{equation*}
  \Var(\hat{\betab}\mid \vb)=(\xb^T \Dv \xb)^{-1}\xb^T \Var(\Dv\yb\mid \vb)\xb(\xb^T \Dv \xb)^{-1} =\sigma^2(\xb^T \Dv \xb)^{-1},
\end{equation*}
since $\Dv\yb\midd \vb\simdist \N_n(\Dv\xb\betab, \sigma^2 \Dv)$. The robust estimation of $\Var(\hat{\betab})$ is then given by $\sigma^2(\xb^T \Dpi \xb)^{-1}$.
Using the same method, we obtain
$\hat{\betab}\approxdist \N_p(\betab, \sigma^2 (\xb^T \Dpi \xb)^{-1})$ and
\begin{equation*}
  \frac{\hat{\beta}_j-\beta_j}{\hat{\sigma}\sqrt{\left[(\xb^T \Dpi \xb)^{-1}\right]_{j,j}}}\approxdist
   t_{\hat{\omega}n - p} \text{ for }j=1,\ldots,p, \,\,\text{ and }\,\, \frac{(\hat{\omega}n - p)\hat{\sigma}^2}{\sigma^2}\approxdist \chi^2_{\hat{\omega}n - p},
\end{equation*}
where $\approxdist$ stands for ``approximately distributed''.
The robust $1-\alpha$ confidence intervals for the parameters are given by
\begin{equation}
\label{eqn:conf-int}
  \beta_j\in \hat{\beta}_j \pm t_{\alpha/2;\hat{\omega}n - p}\, \hat{\sigma}\sqrt{\left[(\xb^T \Dpi \xb)^{-1}\right]_{j,j}}\,\, \text{ and }\,\,
  \frac{(\hat{\omega}n - p)\hat{\sigma}^2}{\chi^2_{\alpha/2;\hat{\omega}n - p}} \le \sigma^2 \le \frac{(\hat{\omega}n - p)\hat{\sigma}^2}{\chi^2_{1-\alpha/2;\hat{\omega}n - p}}.
\end{equation}
The robust coefficient of determination $R^2$ and its adjusted version $\bar{R}^2$ are given by
\begin{equation}
\label{eqn:coef-det}
   R^2 = 1- \frac{(\hat{\omega}n-p)\hat{\sigma}^2}{SSY}\,\,\text{ and }\,\,\bar{R}^2 = 1- \frac{\hat{\sigma}^2}{SSY/(\hat{\omega}n-1)},
\end{equation}
where
\begin{equation}
\label{eqn:ssy}
   SSY= \sum_{i=1}^{n}\hat{\pi}_i \left(y_i-\frac{\sum_{i=1}^{n}\hat{\pi}_i\, y_i}{\sum_{i=1}^{n}\hat{\pi}_i}\right)^2.
\end{equation}
Hypothesis testing and model selection equations can be derived with the same method; see Section~\ref{sec:example} for an example.

\subsection{Outlier Detection}
\label{sec:outlier-detection}

A direct tool for outlier identification is obtained because $\hat{\pi}_i\equiv\pi_{\hat{\omega}}(r_i)$
represents the probability that the observation $y_i$ is a normal observation and thus $1-\pi_{\hat{\omega}}(r_i)$
is the probability that it is an outlier, where  $r_i$ is the (standardized) residual as given in Definition~\ref{def:estimators}.
For observations that have a residual in the central part, that is $|r_i|\le\hat{\tau}$, the probability of being a normal observation is estimated at $\pi_{\hat{\omega}}(r_i)=1$,
where $\hat{\tau}=g^{-1}(\hat{\omega})$ as given in Definition~\ref{def:FLP-contaminated-normal}.
For those with a residual in the outlier region, that is $|r_i|>\hat{\tau}$,
the estimated probability of being an outlier increases gradually from 0 to 1 as $|r_i|$ increases from $\hat{\tau}$ to infinity.
If the probability of being an outlier needs to be converted into a binary decision, it suffices to flag as outliers the observations $i$ such that
$\pi_{\hat{\omega}}(r_i)<0.5$.

Given that the usual method of outlier detection in the literature consists in identifying the observations
with a residual $|r_i|$ larger than a threshold often set to 2.5 (see, e.g., \cite{GerviniYohai2002rewlse}), it can be interesting
to express our criterion $\pi_{\hat{\omega}}(r_i)<0.5$ in the same form.
Considering that $\pi_{\hat{\omega}}(r_i)$ is a decreasing function of $|r_i|$, for $|r_i|>\hat{\tau}$, we have
\begin{equation*}
  \pi_{\hat{\omega}}(r_i)<0.5 \Leftrightarrow |r_i| > \pi^{-1}_{\hat{\omega}}(0.5).
\end{equation*}
Instead of a fixed threshold such as 2.5, we obtain an adaptive cut-off $\pi^{-1}_{\hat{\omega}}(0.5)$ that increases with the estimated proportion of normal
observations $\hat{\omega}$. For example, we obtain $\pi^{-1}_{\hat{\omega}=0}(0.5)=2.466$,
$\pi^{-1}_{\hat{\omega}=0.5}(0.5)=2.516$, $\pi^{-1}_{\hat{\omega}=0.8}(0.5)=2.621$ and $\pi^{-1}_{\hat{\omega}=0.95}(0.5)=2.863$,
which is somewhat consistent with the usual cut-off of 2.5.
The threshold $\pi^{-1}_{\hat{\omega}}(0.5)$ tends to infinity as $\hat{\omega}\rightarrow 1$, meaning no outliers can be identified when $\hat{\omega}=1$, which
is consistent with the fact that outliers have been defined as observations with extreme errors that conflict with the normal assumption.

\section{Example}
\label{sec:example}

A specific dataset is analyzed in this section using N-FLP estimators. The analysis is done from different perspectives that lead to special cases such as location-scale inference,
simple and multiple linear regression, ANOVA and Student's $t$-test.

\begin{figure}[ht]
\begin{center}
\begin{tabular}{cc}
\includegraphics[width=0.49\textwidth]{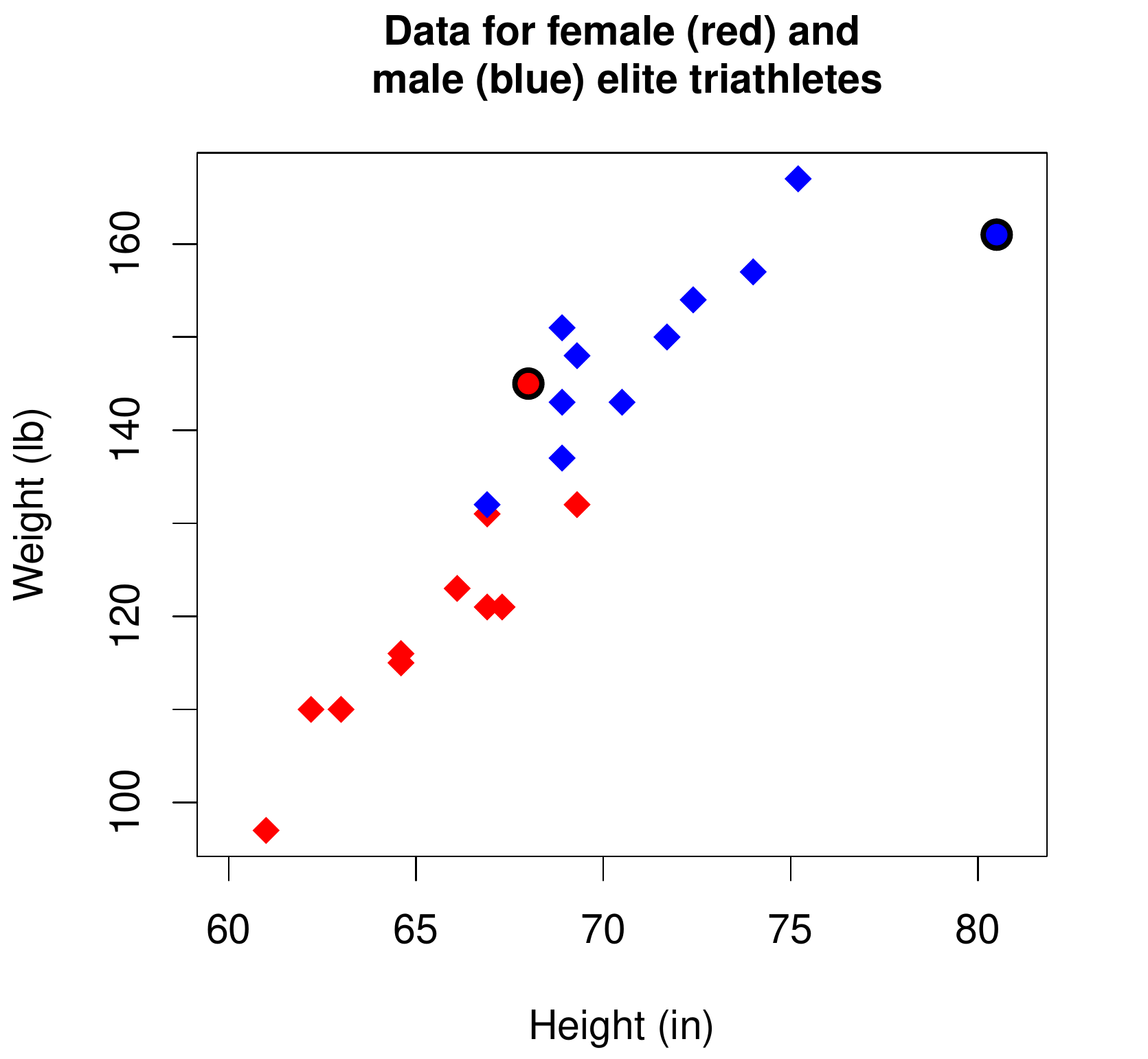}
\end{tabular}
\end{center}
\caption{Data for female (red) and male (blue) elite triathletes.}
\label{fig:data_triathletes}
\end{figure}

A random sample of elite triathletes, 10 females and 10 males, was selected from the International Triathlon Union website, with their weight (in pounds) and height (in inches). The dataset is plotted
in Figure~\ref{fig:data_triathletes}. We added two artificial outliers (the circles in the graph), one for each gender, for a total of $22$ observations.
The dependent variable $y$ is defined as weight and the two explanatory variables $x_2$ and $x_3$ are defined respectively as height and
gender, coded as 0 for female and 1 for male. The detailed dataset and programming code using R software are available in Appendix~D in the supplemental material.

\subsection{Analysis of One Variable Using the Location-Scale Model}
\label{sec:location-scale}

The dataset can be analyzed for one variable at a time, say $y_1,\ldots,y_n$, using the linear regression model with $p=1$, $\beta_1=\mu$ and $\xb_i=1$.
We obtain the location-scale model $y_i\simdist\NFLP(\omega, \mu,\sigma)$, where $\mu\in\re$ and $\sigma>0$ are the location and scale parameters.
The N-FLP estimators of $\mu$ and $\sigma$ can thus be rewritten as
\begin{equation*}
  \hat{\mu}=\frac{\sum_{i=1}^n \hat{\pi}_i\, y_i}{\sum_{i=1}^{n}\hat{\pi}_i} \,\,\text{ and }\,\,
  \hat{\sigma}^2=\frac{1}{(\sum_{i=1}^{n}\hat{\pi}_i-1)}\sum_{i=1}^n \hat{\pi}_i(y_i-\hat{\mu})^2,
\end{equation*}
and  the robust estimation of  $\Var(\hat{\mu})$ is given by $(\hat{\omega}n)^{-1}\hat{\sigma}^2$.

If no outliers are detected, the N-FLP estimators are given by $\hat{\omega}=1$ (with $\hat{\pi}_i=1$ for all observations) and the classical sample mean and standard deviation.
This is the case for the weight, where we obtain $\hat{\mu}_y=120.1$, $\hat{\sigma}_y=12.9$  for females and
$\hat{\mu}_y=149.4$, $\hat{\sigma}_y=10.3$ for males.
This is also the case for females' height, with $\hat{\mu}_{x_2}=65.4$, $\hat{\sigma}_{x_2}=2.6$.

As for the height of the $n=11$ males, the adapted EM algorithm of the N-FLP approach proposes two solutions: the non-robust
$\hat{\omega}=1$, $\hat{\mu}_{x_2}=71.6$, $\hat{\sigma}_{x_2}=3.9$ and the robust
$\hat{\omega}=0.91865$, $\hat{\mu}_{x_2}=70.7723$, $\hat{\sigma}_{x_2}=2.7971$. We opt for the robust solution, as prescribed in the automatic procedure.
The values of $\hat{\pi}_i$ are equal to 1 for all male observations except for the blue circle in the graph which has a value of $\hat{\pi}_i=0.1052$.
 It can thus be flagged as an outlier since the value is smaller than 0.5 and its probability of being an outlier is estimated at $1-\hat{\pi}_i=0.8948$.
Robust $95\%$ confidence intervals  for $\mu_{x_2}$ and $\sigma_{x_2}$ can be computed using equations in~(\ref{eqn:conf-int}).
Given the number of normal observations estimated at
$\hat{\omega}n=\sum_{i=1}^n \hat{\pi}_i=10.1052$ and the quantiles $t_{0.025;9.1052}= 2.25818$, $\chi^2_{0.025;9.1052}=19.1776$
and $\chi^2_{0.975;9.1052}=2.7567$, we obtain
\begin{equation*}
  \mu_{x_2}\in \hat{\mu}_{x_2} \pm t_{0.025;\hat{\omega}n - 1} (\hat{\omega}n)^{-1/2}\,\hat{\sigma}_{x_2} =(68.785, 72.759)\,\text{ and }\,\sigma_{x_2}\in (1.927, 5.083).
\end{equation*}

\subsection{Simple Linear Regression}
\label{sec:simple-linear-regression}

The relationship between weight ($y$) and height ($x_2$) is now analyzed separately for female and male triathletes, using the simple linear regression model
$y_i=\beta_1 + \beta_2 x_{2i} + \varepsilon_i$.
For this special case, the N-FLP estimators can be rewritten as
\begin{equation*}
  \hat{\beta}_1=\frac{\sum_{i=1}^{n}\hat{\pi}_i\, y_i}{\sum_{i=1}^{n}\hat{\pi}_i} - \hat{\beta}_2\, \frac{\sum_{i=1}^{n}\hat{\pi}_i\, x_{2i}}{\sum_{i=1}^{n}\hat{\pi}_i}
  \,\, \text{ and }\,\,\hat{\beta}_2=\frac{SXY}{SSX},\,\text{where}
\end{equation*}
\begin{equation*}
  SSX= \sum_{i=1}^{n}\hat{\pi}_i\!\left(x_{2i}-\frac{\sum_{i=1}^{n}\hat{\pi}_i\, x_{2i}}{\sum_{i=1}^{n}\hat{\pi}_i}\right)^2\!\!,\,
  SXY= \sum_{i=1}^{n}\hat{\pi}_i\!\left(x_{2i}-\frac{\sum_{i=1}^{n}\hat{\pi}_i\, x_{2i}}{\sum_{i=1}^{n}\hat{\pi}_i}\right)\!\left(y_i-\frac{\sum_{i=1}^{n}\hat{\pi}_i\, y_i}{\sum_{i=1}^{n}\hat{\pi}_i}\right)\!.
\end{equation*}
The robust estimations of the variances are given by
\begin{equation*}
  \widehat{\Var}(\hat{\beta}_1)=\frac{\hat{\sigma}^2}{SSX}\, \frac{\sum_{i=1}^{n}\hat{\pi}_i\, x_{2i}^2}{\sum_{i=1}^{n}\hat{\pi}_i}
  \,\, \text{ and }\,\,
  \widehat{\Var}(\hat{\beta}_2)=\frac{\hat{\sigma}^2}{SSX}
\end{equation*}
and the sample correlation coefficient is given by $r_{xy}=SXY/\sqrt{SSX\cdot SSY}$, where $SSY$ is defined in (\ref{eqn:ssy}).

For the $n=11$ female triathletes, the non-robust N-FLP solution, always given by the OLS estimates, is $\hat{\omega}=1$,
$\hat{\beta}_1=-170.3340$, $\hat{\beta}_2=4.4377$ and  $\hat{\sigma}=6.2165$.  However we choose the only robust solution proposed by the algorithm, given by
$\hat{\omega}=0.90985$, $\hat{\beta}_1=-130.6676$, $\hat{\beta}_2=3.8086$ and  $\hat{\sigma}=3.9227$.
The values of $\hat{\pi}_i$ are therefore equal to 1 for all female observations except for the red circle in the graph, which has a value of $\hat{\pi}_i=0.0083$,
meaning it is clearly an outlier. We observe that $\hat{\sigma}$ and $\hat{\beta}_2$ are significantly lower in the robust solution, where the impact of the outlier
is almost wiped out. We find that an increase of 1 inch in height is associated with an average increase of $\hat{\beta}_2=3.8086$ pounds in weight.
The robust sample  correlation coefficient is given by $r_{xy}=0.936$ compared with the non-robust $r_{xy}=0.890$.
Note that the coefficient of determination is simply $R^2=r_{xy}^2$ in this case.

The analysis for the 11 male triathletes gives similar results, with $\hat{\omega}=0.90941$, $\hat{\beta}_1=-96.6101$, $\hat{\beta}_2=3.4640$, $\hat{\sigma}=4.7423$
and $r_{xy}=0.897$.
The blue circle in the graph is still flagged as an outlier, but this time for its large vertical distance from the regression line, with a value of $\hat{\pi}_i=0.0035$.
Equivalently, we can identify as outliers all the observations with a standardized residual $|r_i|$ larger than the adaptive cut-off
$\pi^{-1}_{\hat{\omega}=0.90941}(0.5)=2.7489$. Of course, the blue circle is still the only outlier, with a large residual of $|r_i|=4.4797$.

\subsection{Multiple Linear Regression}
\label{sec:multiple-linear-regression}

We first study the model  $y_i=\beta_1 + \beta_2 x_{2i} + \beta_3 x_{3i} + \beta_4 x_{2i}x_{3i} + \varepsilon_i$ using the whole sample of the $n=22$ triathletes.
Other than the non-robust OLS estimates, only one robust solution is proposed by the algorithm.  The robust model using the N-FLP estimates is then given by
$y_i= -132.5030 + 3.8377\, x_{2i} + 35.2106\, x_{3i} -0.3640\, x_{2i}x_{3i} + \varepsilon_i$, with $\hat{\omega}=0.91112$ and $\hat{\sigma}=4.4113$.
The same two observations as above are flagged as outliers, with probabilities $1-\hat{\pi}_i$ of being an outlier of 0.9540 and 0.9992 respectively
for the red and blue circles in the graph.  We find that an increase of 1 inch in height is associated with an average increase of $\hat{\beta}_2=3.8377$ pounds in weight
for female triathletes ($x_{3i}=0$) and $\hat{\beta}_2+\hat{\beta}_4=3.8377-0.3640=3.4737$ pounds for males ($x_{3i}=1$), similar to what we found in the simple linear
regression analyses. The average difference in weight between a female and male triathlete with the same height $x_{2i}$
is estimated at $\hat{\beta}_3+\hat{\beta}_4 \, x_{2i}=35.2106-0.3640\, x_{2i}$, for example a difference of 10.4586 pounds for a height of 68 inches.

In light of these results, it appears that the association between height and weight is very similar for female and male triathletes
($\hat{\beta}_2\approx \hat{\beta}_2+\hat{\beta}_4$) and it could be reasonable to consider
the simpler model $y_i=\beta_1 + \beta_2 x_{2i} + \beta_3 x_{3i} + \varepsilon_i$ using the hypothesis testing $H_0: \beta_4 = 0$ vs  $H_1: \beta_4 \ne 0$.
To this end, we compute the test statistic $T$ given by $\hat{\beta_4}=-0.3640$ divided by the square root of $\widehat{\Var}(\hat{\beta_4})= \hat{\sigma}^2[(\xb^T \Dpi \xb)^{-1}]_{4,4}=0.6400$ and we obtain $T=-0.4549$. The p-value of $2 P(t_{\hat{\omega}n - p}>|T|)=2 P(t_{16.0447}>0.4549)=0.6553$ does not allow us
to reject $H_0: \beta_4 = 0$ with enough confidence so we choose the smallest model.
It is robustly estimated  by
$y_i= -120.5569 + 3.6537\, x_{2i} + 10.4821\, x_{3i}  + \varepsilon_i$, with $\hat{\omega}=0.90655$ and $\hat{\sigma}=4.2158$, and the same two outliers are identified.
We now find that an increase of 1 inch in height is associated with an average increase of $\hat{\beta}_2=3.6537$ pounds in weight for any gender,
and that the average difference in weight between a female and male triathlete, for any given height,
is estimated at 10.4821 pounds. The robust coefficient of determination $R^2$ and its adjusted version $\bar{R}^2$ can be computed
using equations in~(\ref{eqn:coef-det}) and we find $R^2 = .954$ (vs 0.875 for OLS) and $\bar{R}^2 =0.949$ (vs 0.862 for OLS).

\subsection{ANOVA and Student's $t$-Test}
\label{sec:ANOVA}

We now analyze the body mass index (BMI), a well-known measure of weight (in kg) per unit of squared height (in $m^2$), also defined as
 $y= \text{weight (lb)} \div \text{height}^2\,(\text{in}^2) \times 703.07$.
Precisely, we want to assess if there is a difference between the average BMI of female and male triathletes
 using the one-way analysis of variance (ANOVA) with unbalanced data,
which is also the independent two-sample $t$-test with equal variance when there are only two groups as in our case.
The fixed effect model is considered, which means that an observation is outlying relative to its group only. Inter-group outliers are not considered, as is
the case with mixed models. Although this is a particular case of linear regression, it is worthwhile to write the ANOVA model in its usual form.

We are given a random sample $y_{i,j}$, for $i=1,\ldots,n_j$ and $j=1,\ldots,J$, having the model $y_{i,j}=\mu_j +\varepsilon_{i,j}$, with
$\varepsilon_{i,j}\simdist \NFLP(\omega, 0,\sigma)$, or equivalently $y_{i,j}\simdist \NFLP(\omega, \mu_j,\sigma).$
There are $J$ treatment groups or populations ($J=2$ for the $t$-test) and
$n_j$ experimental units in the group $j$, for a grand total of $n=\sum_{j=1}^{J}n_j$ observations. The location of the population $j$
is $\mu_j\in\re$ and $\sigma>0$ is the same scale parameter for all $n$ observations.
If $\omega=1$, we obtain the classical ANOVA and $t$-test models with Gaussian distributions.

The N-FLP estimators can be rewritten as
\begin{equation*}
  \hat{\omega}=\frac{1}{n}\sum_{j=1}^{J}\sum_{i=1}^{n_j} \hat{\pi}_{i,j},\,
  \hat{\mu}_j=\frac{\sum_{i=1}^{n_j} \hat{\pi}_{i,j}\, y_{i,j}}{\sum_{i=1}^{n_j} \hat{\pi}_{i,j}}\text{ and }\hat{\sigma}^2
  =\frac{1}{(\sum_{j=1}^{J}\!\sum_{i=1}^{n_j} \hat{\pi}_{i,j}-J)}\sum_{j=1}^{J}\sum_{i=1}^{n_j} \hat{\pi}_{i,j}(y_{i,j}-\hat{\mu}_j)^2,
\end{equation*}
\begin{equation*}
 \text{with }\hat{\pi}_{i,j}\equiv\pi_{\hat{\omega}}(r_{i,j})=\frac{\hat{\omega} f_{\N}(r_{i,j}\mid 0, 1)} {f_{\NFLP}(r_{i,j}\mid \hat{\omega}, 0, 1)},\,\,\text{ where }\,\,r_{i,j}=\frac{y_{i,j}- \hat{\mu}_j}{\hat{\sigma}}.
\end{equation*}

The test $H_0 : \mu_1 = \ldots \mu_J$ is done using the statistic
\begin{equation*}
  F=\frac{SS_{tr}/(J-1)}{\hat{\sigma}^2}\approxdist F_{J-1,\hat{\omega}n-J}\text{ under }H_0,
\end{equation*}
where
\begin{equation*}
  SS_{tr}=\sum_{j=1}^{J}\left(\hat{\mu}_j - \frac{\sum_{j=1}^{J}\sum_{i=1}^{n_j}  \hat{\pi}_{i,j}\, y_{i,j}}{\sum_{j=1}^{J}\sum_{i=1}^{n_j}  \hat{\pi}_{i,j}}\right)^2
   \sum_{i=1}^{n_j}\hat{\pi}_{i,j}
\end{equation*}
is the sum of squares for the treatment.
The statistic for the $t$-test $H_0: \mu_1=\mu_2$ vs $H_1: \mu_1\ne\mu_2$ can also be written as
\begin{equation*}
  T=\frac{\hat{\mu}_2-\hat{\mu}_1}{\hat{\sigma}\sqrt{(\sum_{i=1}^{n_1} \hat{\pi}_{i,1})^{-1}+(\sum_{i=1}^{n_2} \hat{\pi}_{i,2})^{-1}}}\approxdist t_{\hat{\omega}n-2}\text{ under }H_0.
\end{equation*}

Now, for the analysis of BMI, the $J=2$ groups are represented by female (group 1) and male (group 2) triathletes, with $n_1=n_2=11$ observations in each group.
The robust N-FLP estimates are given by $\hat{\omega}=0.90407$,
$\hat{\mu}_1=19.4268$, $\hat{\mu}_2=20.8346$ and $\hat{\sigma}=0.6635$. The statistic for the  $t$-test is evaluated at $T=4.7310$
given that $\sum_{i=1}^{n_1} \hat{\pi}_{i,1}=10.02335$ and $\sum_{i=1}^{n_2} \hat{\pi}_{i,2}=9.86615$.
The p-value of $2 P(t_{\hat{\omega}n - 2}>|T|)=2 P(t_{17.8895}>4.7310)=0.00017$ allows us to conclude that the average BMI for female and male triathletes is
significantly different.  Note that the non-robust solution, given by the classical $t$-test, concludes instead that
$\hat{\mu}_1=19.6594$, $\hat{\mu}_2=20.5472$ and that their difference is not significant at a level of 0.05, with a p-value of 0.0758.

\section{Monte Carlo Simulation}
\label{sec:MonteCarlo}

In this section, the performance of eight estimators is compared in a Monte Carlo simulation. In addition to the N-FLP estimators proposed in this paper
and the OLS estimators considered as the benchmark,
we include the REWLS, MM, M (using Tukey's bisquare $\psi$ function), S, LMS and LTS estimators, as \cite{yuyao2017review} did in their review.
The detailed programming code using R software is provided in Appendix~D in the supplemental material.

The performance of the given estimators $\hat{\betab}$ and $\hat{\sigma}$ is measured respectively by $E[D_{\hat{\betab}}(\betab)]$ and
$E[D_{\hat{\sigma}}(\sigma)]$, where the expectations are taken with respect to the random sample $(\xb_i, y_i)$, $i=1,\ldots,n$
and $\betab,\sigma$ are the true parameters of the linear regression model.
The affine and scale invariant distances $D_{\hat{\betab}}(\betab)$ and $D_{\hat{\sigma}}(\sigma)$ are defined, for a given sample, as
\begin{equation*}
  D_{\hat{\betab}}(\betab)=n^{-1/2}\sqrt{(\hat{\betab}-\betab)^T \sigma^{-2}(\xb^T \xb)(\hat{\betab}-\betab)}\hspace{5mm}\text{ and }\hspace{5mm}
  D_{\hat{\sigma}}(\sigma)=\abs{\log(\hat{\sigma}/\sigma)},
\end{equation*}
as proposed by \cite{GerviniYohai2002rewlse}.
The distance $D_{\hat{\sigma}}(\sigma)$ is meant to account for both the explosion and implosion of the scale estimator.
We observe that $n^{1/2}D_{\hat{\betab}}(\betab)$ corresponds to the Mahalanobis distance between
$\hat{\betab}$ and $\betab$. The distance $D_{\hat{\betab}}(\betab)$ can also be rewritten (the proof is given in Appendix~B in the supplemental material) as
\begin{equation*}
  D_{\hat{\betab}}(\betab)=\sigma^{-1}\sqrt{\frac{1}{n}\sum_{i=1}^n \left(\xb_i^T\hat{\betab} - \xb_i^T\betab\right)^2},
\end{equation*}
which can be interpreted as a distance between the estimated hyperplan $y_i = \xb_i^T\hat{\betab}$
and the true one $y_i = \xb_i^T\betab$.
Note that if $p=1$, we have $D_{\hat{\betab}}(\betab)=\sigma^{-1}|\hat{\beta}_1-\beta_1|$ and if the $p>1$  explanatory variables
are uncorrelated and standardized, we obtain $n^{-1}\xb^T \xb = \I_p$  and
$D_{\hat{\betab}}(\betab)=\sigma^{-1}\sqrt{(\hat{\betab}-\betab)^T(\hat{\betab}-\betab)}$,
a simplified distance between $\hat{\betab}$ and $\betab$ sometimes used in the literature.
The expectation of these distances, with respect to the random sample, is estimated using Monte Carlo simulations by calculating the average
of the simulated distances.

The core model studied in this section is the simple linear regression model ($p=2$) given by $y=\beta_1+\beta_2\, x+\varepsilon$, where
$\varepsilon/\sigma\simdist \N(0,1)$.
Models with $p=1$ and $p=5$ have also been considered, but conclusions were very similar.
Without loss of generality, the true parameters are set to  $\beta_1=\beta_2=0$, $\sigma=1$ and the explanatory variable
is generated independently as $x\simdist \N(0,1)$. Said otherwise, a
random sample $(x_1,y_1),\ldots, (x_n,y_n)$ is simulated from a bivariate normal distribution $\N_2(0, \I_2)$.
In Section~\ref{sec:outliers}, different fractions of the sample are modified to
include outliers, but first the core model without contamination is considered as such in Section~\ref{sec:efficiency} in order to study the performance of the estimators
in the absence of outliers.

\subsection{Efficiency under the Uncontaminated Model}
\label{sec:efficiency}

Robust estimators always come at a cost that translates into deteriorated performance in the absence of outliers, relative to the OLS estimates.
This cost, for the given estimators $\hat{\betab}$ and $\hat{\sigma}$, can be measured by the relative efficiency (RE) that we define as
\begin{equation}
\label{eqn:re}
\text{RE}_{\hat{\betab}} = (E[D_{\hat{\betab}_0}(\betab)]\big/E[D_{\hat{\betab}}(\betab)])^2
\text{ and }\text{RE}_{\hat{\sigma}} = \left(E[D_{\hat{\sigma}_0}(\sigma)]\big/E[D_{\hat{\sigma}}(\sigma)]\right)^2,
\end{equation}
where $\hat{\betab}_0$ and $\hat{\sigma}_0$ represent respectively the OLS estimates of $\betab$ and $\sigma$.
Note that the squared distances are used to obtain similar results to relative efficiency computed with mean squared errors, as is often done in the literature.

To estimate the relative efficiencies, 100,000 random samples were simulated under the uncontaminated core model using $(x_1,y_1),\ldots, (x_n,y_n)\simdist \N_2(0, \I_2)$,
for each of the sample sizes $n=50$, 100, 200 and 500. For each generated sample,
$\hat{\betab}$, $\hat{\sigma}$, $D_{\hat{\betab}}(\betab)$ and $D_{\hat{\sigma}}(\sigma)$ were computed for each of the eight estimators in the study.
The expected distances $E[D_{\hat{\betab}}(\betab)]$ and $E[D_{\hat{\sigma}}(\sigma)]$ were estimated by the average of the corresponding 100,000 simulated distances,
for each estimator and sample size, and finally the RE were computed using~(\ref{eqn:re}).
The results are given in Table~\ref{table_eff}. The maximum value (the best performance) for each column of the table appears in red.

\begin{table}[b]
\footnotesize
\begin{center}
\begin{tabular}{|l|cccc|cccc|}
\hline
&& & \hspace{-9mm}$\hat{\betab}$ & & & & \hspace{-9mm}$\hat{\sigma}$ & \cr
\hline
\textbf{Sample size:} & \textbf{50} & \textbf{100} & \textbf{200} & \textbf{500}&  \textbf{50} & \textbf{100} & \textbf{200} & \textbf{500} \cr
\hline
OLS      &   1.000   &     1.000   &     1.000   &     1.000   &     1.000   &      1.000    &     1.000    &     1.000 \cr
N-FLP     &   \textbf{\textcolor{red}{0.982}}   &     \textbf{\textcolor{red}{0.991}}   &     \textbf{\textcolor{red}{0.995}}   &     \textbf{\textcolor{red}{0.998}}   &
            \textbf{\textcolor{red}{0.882}}   &      \textbf{\textcolor{red}{0.924}}    &     \textbf{\textcolor{red}{0.944}}    &     \textbf{\textcolor{red}{0.964}} \cr
MM       &   0.941   &     0.946   &     0.946   &     0.948   &     0.524   &      0.537    &     0.537    &     0.541 \cr
M        &   0.937   &     0.944   &     0.946   &     0.948   &     0.362   &      0.367    &     0.366    &     0.369 \cr
REWLS    &   0.862   &     0.896   &     0.912   &     0.924   &     0.487   &      0.504    &     0.506    &     0.512 \cr
S        &   0.307   &     0.291   &     0.288   &     0.285   &     0.524   &      0.538    &     0.537    &     0.541 \cr
LMS      &   0.176   &     0.149   &     0.126   &     0.096   &     0.167   &      0.229    &     0.244    &     0.189 \cr
LTS      &   0.168   &     0.134   &     0.111   &     0.091   &     0.285   &      0.341    &     0.341    &     0.253 \cr
 \hline
 \end{tabular}
\end{center}
 \vspace{-3mm}
 \caption{Relative efficiency for the uncontaminated model. \label{table_eff}}
\normalsize
\end{table}

Several conclusions can be drawn. First, N-FLP estimators clearly have the highest RE among the robust alternatives,
both for $\hat{\betab}$ and $\hat{\sigma}$ and for all sample sizes in the study (see the red line in Table~\ref{table_eff}). Second, the efficiency of the N-FLP estimator
of $\betab$ is very close to that of the OLS, with RE increasing with the sample size from 0.982 to 0.998. In comparison, the MM, M and REWLS estimators of $\betab$ reach their maximum
at $n=500$ with RE of respectively 0.948, 0.948 and 0.924, which is also very good. Third, the performance of the N-FLP estimator of $\sigma$ is
nearly twice as strong as that of the competition.
The RE of the robust alternatives range from 0.167 to 0.541, compared to 0.882 ($n=50$) to 0.964 ($n=500$) for the N-FLP estimator.
Fourth, the S, LMS and LTS estimators have poor relative efficiencies both for $\hat{\betab}$ and $\hat{\sigma}$, which
disqualifies them from the outset if robust and efficient estimators are wanted.

The high efficiency of the N-FLP estimators when the errors are Gaussian can be largely explained by the fact that
approximately 90\% of the simulated samples produced only the non-robust N-FLP solution, that is the OLS estimates, which means no cost at all most of the time
both for $\betab$ and $\sigma$. This nice feature should be appreciated by practitioners mainly familiar with OLS.
Among the robust alternatives in our study, only the REWLS estimator of $\betab$ has this property, and only for a small
fraction of samples. The probabilities that the N-FLP (for $\betab$ and $\sigma$) and REWLS (for $\betab$ only) estimates are
identical to those of the OLS have been estimated by 100,000 Monte Carlo simulations for sample sizes $n$  of 50, 100 and 200,
for the uncontaminated core models with $p=2$ and $p=5$. The results are given in
Table~\ref{table_prop}. The probabilities for the REWLS estimator are relatively small and rapidly decrease with both $n$ and $p$, from 0.255 to 0.011.
In contrast, the probabilities for the N-FLP range from 0.881 to 0.920, which represents a huge improvement.

\begin{table}[b]
\footnotesize
\begin{center}
\begin{tabular}{|l|ccc|ccc|}
\hline
&&$p=2$ &  & &$p=5$   & \cr
\hline
\textbf{Sample size:} & \textbf{50} & \textbf{100} & \textbf{200} & \textbf{50} & \textbf{100} & \textbf{200}  \cr
\hline
N-FLP            & \textbf{\textcolor{red}{0.916}}        & \textbf{\textcolor{red}{0.898}}        & \textbf{\textcolor{red}{0.881}}   & \textbf{\textcolor{red}{0.920}}   & \textbf{\textcolor{red}{0.909}}         & \textbf{\textcolor{red}{0.887}}\cr
REWLS       & 0.255      & 0.117        & 0.027    &  0.097  & 0.046        & 0.011  \cr
 \hline
 \end{tabular}
\end{center}
 \vspace{-3mm}
 \caption{Probabilities, under the uncontaminated model, that the N-FLP (for $\betab$ and $\sigma$) and REWLS (for $\betab$ only) estimates are
   identical to those of the OLS. \label{table_prop}}
\normalsize
\end{table}

\subsection{Efficiency in the Presence of Outliers}
\label{sec:outliers}

The performance of the eight estimators is now studied when the sample contains outliers.
The sample size is set to $n=50$. Other sample sizes such as $n=100$ were considered, but results were similar.
A large range of contamination scenarios is investigated. Each scenario starts with the uncontaminated core model (with $p=2$), and
$k\%$ of the observations in each sample are replaced by identical outliers $(x_0, y_0)$.
We set $x_0=1$ and $x_0=10$ in order to study the influence of low- and high-leverage outliers, as \cite{GerviniYohai2002rewlse} did.
For each of case, the percentage of contamination is set to $k\in\{2, 4, 10, 20, 30\}$.
A scenario of $k=40$  is also added for the low-leverage case, for a total of 11 combinations of $(x_0, k)$.

The values of $y_0$ remain to be selected to complete the scenarios. Instead of choosing a single and arbitrary value for each combination,
we go a step further by choosing, for all intents and purposes, all values of $y_0$. To achieve this,
a range of $[0, y_{max}]$ is carefully selected for each of the 11 combinations $(x_0, k)$, such that for any value $y_0$
beyond $y_{max}$, the influence of the outliers $(x_0, y_0)$ on the robust estimators becomes stable.
The interval $[0, y_{max}]$ is then evenly split into 31 values of $y_0$, which is sufficient to obtain a smooth pattern.
This represents 31 different scenarios for each of the 11 combinations, for a total of 341 scenarios or combinations of $(x_0, k, y_0)$.
This should allow us to obtain a clear picture of the situation. For each of these 341 scenarios,
the distances $E[D_{\hat{\betab}}(\betab)]$ and $E[D_{\hat{\sigma}}(\sigma)]$ are estimated for each estimator in the study
using 10,000 Monte Carlo simulations.

The results are first presented in detail in the form of graphs in Appendix~C in the supplemental material.
For each of the 11 combinations $(x_0, k)$, a distinct graph for the estimation of $\betab$ and $\sigma$ is presented, for a total of 22 graphs.
The values of $y_0$ lie in the x-axis,
and the distances $E[D_{\hat{\betab}}(\betab)]$ or $E[D_{\hat{\sigma}}(\sigma)]$ lie in the y-axis.
Therefore, each line in a graph represents the performance of a given estimator for a scenario $(x_0, k, y_0)$ with any value of $y_0\in[0, y_{max}]$.
Note that the upper bound of the y-axis was chosen such that the OLS line finishes at the top-right corner in order to consistently visualize the performance
of each estimator relative to the OLS.

The results are also summarized in  Table~\ref{table_low} (low-leverage outliers) and Table~\ref{table_high} (high-leverage outliers) by calculating, for each line
of the graphs,
the average of the distances computed at the 31 different values of $y_0$ in the interval $[0, y_{max}]$. This can be interpreted as a measure of the distance of a line from the
horizontal line located at the origin. The smaller the better. Each column of the tables corresponds to one of the 22 graphs.
The best performance for each column appears in red.

\begin{table}[h]
\footnotesize
\begin{center}
\begin{tabular}{|l|cc|cc|cc|cc|cc|cc|}
\hline
 & &\hspace{-9mm}\textbf{2\%}& &\hspace{-9mm}\textbf{4\%} & &\hspace{-9mm}\textbf{10\%} & &\hspace{-9mm}\textbf{20\%} & &\hspace{-9mm}\textbf{30\%} & &\hspace{-9mm}\textbf{40\%} \cr
& $\hat{\betab}$ & $\hat{\sigma}$ & $\hat{\betab}$ & $\hat{\sigma}$ & $\hat{\betab}$ & $\hat{\sigma}$ & $\hat{\betab}$ & $\hat{\sigma}$ & $\hat{\betab}$ & $\hat{\sigma}$ & $\hat{\betab}$ & $\hat{\sigma}$ \cr
\hline
\hspace{-1.2mm}N-FLP	&	\hspace{-1.1mm}\textbf{\textcolor{red}{0.184}}	&	\hspace{-1.1mm}\textbf{\textcolor{red}{0.095}}	&	0.195	&	\hspace{-1.1mm}\textbf{\textcolor{red}{0.104}}	&	0.253	&	\hspace{-1.1mm}\textbf{\textcolor{red}{0.135}}	&	\hspace{-1.1mm}\textbf{\textcolor{red}{0.318}}	&	\hspace{-1.1mm}\textbf{\textcolor{red}{0.159}}	&	
   \hspace{-1.1mm}\textbf{\textcolor{red}{0.490}}	&	  \hspace{-1.1mm}\textbf{\textcolor{red}{0.222}}	&	\hspace{-1.1mm}\textbf{\textcolor{red}{1.590}}	&	\hspace{-1.1mm}\textbf{\textcolor{red}{0.437}}	\cr
\hspace{-2.6mm} REWLS\!\!	&	0.193	&	0.116	&	0.197	&	0.117	&	\hspace{-1.1mm}\textbf{\textcolor{red}{0.237}}	&	0.149	&	0.325	&	0.285	&	0.639	&	0.486	&	2.435	&	0.817	\cr
\hspace{-1.2mm}MM	&	0.185	&	0.113	&	\hspace{-1.1mm}\textbf{\textcolor{red}{0.192}} 	&	0.116	&	0.248	&	0.154	&	0.405	&	0.301	&	0.929	&	0.512	&	2.689	&	0.867	\cr
\hspace{-1.2mm}S	&	0.318	&	0.113	&	0.316	&	0.116	&	0.331	&	0.155	&	0.405	&	0.301	&	0.793	&	0.512	&	2.718	&	0.868	\cr
\hspace{-1.2mm}LMS	&	0.421	&	0.192	&	0.425	&	0.182	&	0.451	&	0.176	&	0.547	&	0.195	&	1.121	&	0.303	&	5.900	&	0.524	\cr
\hspace{-1.2mm}LTS	&	0.431	&	0.149	&	0.435	&	0.144	&	0.461	&	0.151	&	0.595	&	0.188	&	1.481	&	0.330	&	7.274	&	0.668	\cr
\hspace{-1.2mm}M	&	0.186	&	0.135	&	0.192	&	0.136	&	0.247	&	0.164	&	0.589	&	0.401	&	2.854	&	1.349	&	\!\!\!10.584	&	2.262	\cr
\hspace{-1.2mm}OLS	&	0.204	&	0.155	&	0.278	&	0.248	&	0.570	&	0.414	&	1.579	&	0.807	&	2.829	&	1.034	&	\!\!\!10.114	&	1.930	\cr
 \hline
 \end{tabular}
\end{center}
 \vspace{-3mm}
 \caption{Average distances for low-leverage scenarios. The minimum value for each column appears in red. \label{table_low}}
\normalsize
\end{table}

\begin{table}[h]
\footnotesize
\begin{center}
\begin{tabular}{|l|cc|cc|cc|cc|cc|}
\hline
 & &\hspace{-9mm}\textbf{2\%}& &\hspace{-9mm}\textbf{4\%} & &\hspace{-9mm}\textbf{10\%} & &\hspace{-9mm}\textbf{20\%} & &\hspace{-9mm}\textbf{30\%}  \cr
& $\hat{\betab}$ & $\hat{\sigma}$ & $\hat{\betab}$ & $\hat{\sigma}$ & $\hat{\betab}$ & $\hat{\sigma}$ & $\hat{\betab}$ & $\hat{\sigma}$ & $\hat{\betab}$ & $\hat{\sigma}$ \cr
\hline
N-FLP	&	\textbf{\textcolor{red}{0.285}}	&	\textbf{\textcolor{red}{0.093}}	&	\textbf{\textcolor{red}{0.367}}	&	\textbf{\textcolor{red}{0.096}}	&	\textbf{\textcolor{red}{0.512}}	&	\textbf{\textcolor{red}{0.108}}	&	\textbf{\textcolor{red}{0.718}}	&	\textbf{\textcolor{red}{0.126}}	&	\textbf{\textcolor{red}{1.290}}	&	\textbf{\textcolor{red}{0.170}} \cr
REWLS	&	0.298	&	0.114	&	0.426	&	0.116	&	0.828	&	0.146	&	2.088	&	0.265	&	4.857	&	0.461\cr
MM	&	0.308	&	0.111	&	0.444	&	0.115	&	0.838	&	0.150	&	2.148	&	0.276	&	5.097	&	0.482\cr
S	&	0.453	&	0.111	&	0.590	&	0.115	&	1.008	&	0.150	&	2.282	&	0.276	&	5.122	&	0.482\cr
LMS	&	0.586	&	0.193	&	0.726	&	0.186	&	1.188	&	0.184	&	2.632	&	0.228	&	6.562	&	0.331\cr
LTS	&	0.602	&	0.150	&	0.764	&	0.149	&	1.333	&	0.163	&	3.441	&	0.251	&	9.817	&	0.430\cr
M	&	0.363	&	0.138	&	1.208	&	0.223	&	3.915	&	0.395	&	8.966	&	0.525	&	21.992	&	0.788\cr
OLS	&	0.846	&	0.184	&	1.368	&	0.218	&	3.831	&	0.419	&	8.826	&	0.651	&	21.698	&	1.094\cr
 \hline
 \end{tabular}
\end{center}
 \vspace{-3mm}
 \caption{Average distances for high-leverage scenarios. The minimum value for each column appears in red. \label{table_high}}
\normalsize
\end{table}

Several conclusions can be drawn. First, the N-FLP estimators clearly have the best performance.
Indeed, they have the lowest average distances (the red numbers in the tables) for 9 of 11 scenarios for the estimation of $\betab$
and for the 11 scenarios for $\sigma$. In particular, the domination of the N-FLP estimators is unequivocal for all the high-leverage scenarios (see Table~\ref{table_high})
and for the low-leverage scenarios with a high level of contamination (30\% and 40\%),
both for $\hat{\betab}$ and $\hat{\sigma}$.
Second, the situation is not so clear for the estimation of $\betab$ for the low-leverage scenarios with a lower level of contamination (2\% to 20\%).
However, it appears that the N-FLP, REWLS, MM and M estimators emerge as the best in this situation, where a slight compromise must generally be made between
efficiency in the presence or the absence of outliers.
Third, we can observe in the graphs that good performance of an estimator is reached when the distance
first increases with $y_0$ until a certain threshold and then gradually decreases at a level where the influence of the outliers vanishes.
Practically all the good robust estimators exhibit this behavior for the estimation of $\betab$, but
the N-FLP estimator is the only one for the estimation of $\sigma$, which largely explains its success.

\section{Conclusion}
\label{sec:conclusion}

The objective of this paper was to propose an original, better method for the estimation of a linear regression model with normal errors that may contain outliers.
Our approach first consisted in broadening the classical normal distribution of the errors to
a mixture of the normal and the FLP, an original distribution that we designed such that the outlier region and the tail's behavior are set automatically, based on the proportion of outliers given by the mixture weights. The second step was to propose an original method of estimation for the parameters of the N-FLP mixture, which is
essentially an EM algorithm adapted to our context.

In Monte Carlo simulations where eight of the best and most popular estimators were studied,
 N-FLP estimators clearly had the best performance both in the presence and the absence of outliers, for both the estimation of the regression coefficients
 and the scale parameter. In particular,
 the efficiency of N-FLP estimators when the errors are normally distributed is very close to that of OLS for the regression coefficients and practically
  twice better than the competition for the estimation of the scale
parameter. Furthermore, the N-FLP approach generates exactly the same results as the OLS for approximately 90\%  of the uncontaminated samples,
a nice feature for practitioners mainly used to OLS.
At the same time, efficiency is also greatly improved in the presence of outliers compared with the best alternatives.
No compromise is required: the best efficiencies are obtained both in the presence and the absence of outliers.

Furthermore, we obtained explicit and interpretable expressions for the N-FLP estimators, transparency that is generally appreciated by practitioners.
The procedure can be used routinely where the only requirement made to the statistician is to provide the dataset.
Outlier identification is direct and efficient, where the probability of being an outlier is provided for each observation.
Finally, we showed through an example how N-FLP estimators can be used for a complete inference,
including confidence intervals, hypothesis testing and model selection, for different special cases such as the analysis of one variable
using location-scale inference,
simple and multiple linear regressions, Student's $t$-test and ANOVA.

\section*{Supplemental Material}
\label{sec:supplemental-material}

The supplementary material contains two files:
\begin{verbatim}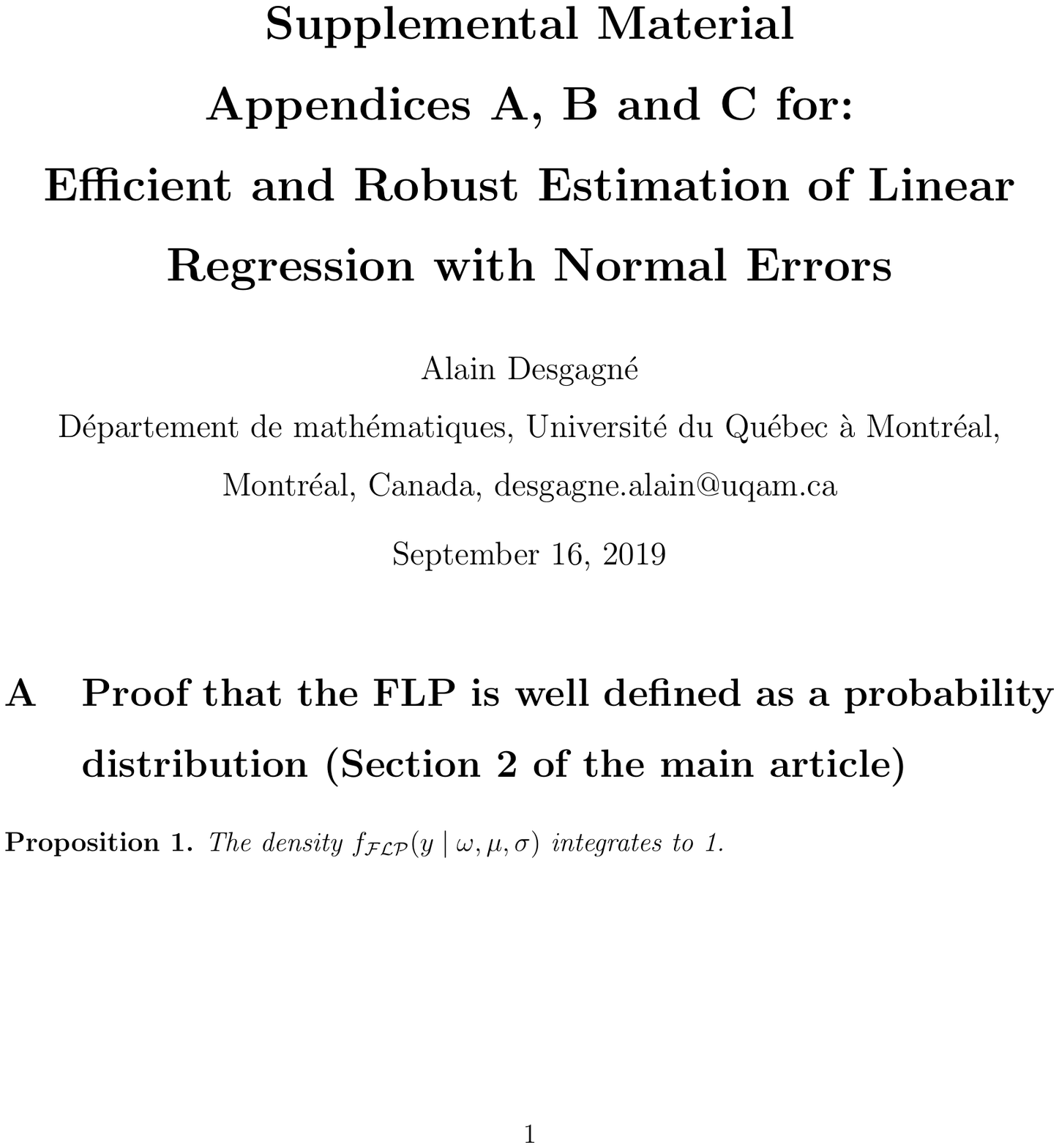\end{verbatim}
\begin{verbatim}Supplemental_Material_AppD_Efficient_and_Robust_Linear_Regression.R\end{verbatim}
In the first file, proofs are given in Appendices A and B and the graphs from the Monte Carlo simulation are given in Appendix C. In the second file,
 the programming code using R software and the dataset for the example are given in Appendix D.

\bibliographystyle{ba}
\bibliography{Efficient_and_Robust_Estimation_of_Linear_Regression_with_Normal_Errors}

\end{document}